\documentstyle[12pt,supercite]{article}

\newcommand{\BF}[1]{\mbox{\boldmath $#1$}}

\def\abstracts#1#2#3{{
        \centering{\begin{minipage}{4.62in}\baselineskip=13pt
        \small
        \centerline{\bf Abstract}
        \vspace*{0.2cm}             
        \parindent=0pt #1\par
        \parindent=18pt #2\par
        \parindent=15pt #3
        \end{minipage} }\par}}
\begin{document}
\vspace*{-2cm}
\hfill \parbox{3cm}{KOMA-96-17\\
                    August 1996 }\\
\vspace*{2cm}
\centerline{\LARGE \bf 
                         Logarithmic Corrections in 
           }\\[0.3cm]
\centerline{\LARGE \bf 
                         the 2D XY Model
           }\\[0.8cm]
%
\vspace*{0.2cm}
\centerline{\large {\em Wolfhard Janke\/}}\\[0.4cm]
\centerline{\large {\small Institut f\"ur Physik,
                    Johannes Gutenberg-Universit\"at Mainz}}
\centerline{    {\small Staudinger Weg 7, 55099 Mainz, Germany }}\\[0.5cm]
\abstracts{}{
Using two sets of high-precision Monte Carlo data for the two-dimensional
XY model in the Villain formulation on square $L \times L$ lattices, the 
scaling behavior of the susceptibility $\chi$ and correlation length $\xi$ 
at the Kosterlitz-Thouless phase transition is analyzed with emphasis on 
multiplicative logarithmic corrections $(\ln L)^{-2r}$ in the finite-size
scaling region and $(\ln \xi)^{-2r}$ in the high-temperature phase near
criticality, respectively. By analyzing the susceptibility at criticality
on lattices of size up to $512^2$ we obtain $r = -0.0270(10)$, in agreement
with recent work of Kenna and Irving on the the finite-size scaling of
Lee-Yang zeros in the cosine formulation of the XY model. By studying
susceptibilities and correlation lengths up to $\xi \approx 140$ in the
high-temperature phase, however, we arrive at quite a different estimate of 
$r = 0.0560(17)$, which is in good agreement with recent analyses of 
thermodynamic Monte Carlo data and high-temperature series expansions 
of the cosine formulation. 
}{}
\vspace*{1.5cm}
%
\thispagestyle{empty}
\newpage
\pagenumbering{arabic}
%
                     \section{Introduction}
%
Ever since the seminal work of Kosterlitz and Thouless (KT) in 
1973 \cite{kos73,kos74}, the two-dimensional (2D) XY model has been 
the subject of extensive experimental, analytical and numerical 
investigations \cite{reviews}. Physically the interest in this model 
arises from studies of layers of superconducting materials and films of
liquid helium \cite{liqhel}, Josephson-junction arrays \cite{joseph},
and some magnetic systems \cite{magsys}. Theoretically the peculiar behavior
of the KT phase transition, which is believed to be driven by the unbinding
of defect pairs, has attracted much interest. Despite all these efforts, 
however, the details of the phase transition are not yet fully understood. 

In a recent Monte Carlo (MC) simulation study of Lee-Yang partition 
function zeros, Kenna and Irving \cite{kenna96,kenna_plb} raised again the
question of logarithmic corrections \cite{kos74,ami80} to the leading 
finite-size scaling (FSS) scaling behavior. If the linear lattice size is 
denoted by $L$ and the multiplicative logarithmic corrections are 
parametrized as 
$(\ln L)^{-2r}$, their numerical result is $r = -0.02(1)$, while the standard
KT theory would predict quite a different exponent of 
$r = -1/16 = -0.0625$ \cite{kos74,ami80}. 
Moreover, by reanalyzing ``thermodynamic'' MC data of Refs.~\cite{wolff,gupta}
obtained on lattices with $L > 7 \xi$, where $\xi$ is the correlation length, 
Patrascioiu and Seiler \cite{seiler} obtained an estimate of $r = 0.077(46)$, 
and by analyzing long high-temperature series expansions, 
Campostrini {\em et al.\/} \cite{campostrini} also arrived at positive 
values in the range
$r = 0.042(5)$ to $r = 0.05(2)$, depending on the quantity considered.
While the estimates of the latter two groups are consistent with each other,
they are incompatible with the FSS result of Kenna and Irving,
which, on the other hand, is somewhat ``closer'' to the theoretical 
prediction.

All numerical estimates quoted above were obtained in the cosine formulation
of the XY model. The purpose of this note is to add further evidence in one or
the other direction by analyzing the logarithmic corrections in the Villain 
formulation \cite{vil75} of the XY model, which is actually (sometimes 
implicitly) the starting point of most if not all theoretical investigations. 
%
                     \section{Scaling predictions}
%
In the Villain XY model \cite{vil75} the Boltzmann factor of the cosine 
formulation, $B_{\rm cos} = \prod_{\BF{x},i} \exp\left[ \beta_{\rm cos} 
\cos(\nabla_i \theta(\BF{x})) \right]$, is replaced by the periodic Gaussian
\begin{equation}
B = \prod_{\BF{x},i}
\sum_{n=-\infty}^\infty \exp
\left[ -\frac{\beta}{2} (\nabla_i \theta - 2\pi n)^2 \right],
\label{eq:B_vil}
\end{equation}
where $\beta$ is the inverse temperature in natural units, and
$\nabla_i \theta = \theta(\BF{x}+\BF{i}) - \theta(\BF{x})$ are lattice
gradients. A discussion of the relation between the two formulations as well as
numerical comparisons can be found in Refs.~\cite{vil75,vil_app}. 

The two-point correlation function ($\vec{s} = (\cos(\theta),\sin(\theta)$)
\begin{equation}
G(\BF{x}) \equiv \langle \vec{s}(\BF{x}) \cdot \vec{s}(\BF{0}) \rangle = 
                 \langle \cos(\theta(\BF{x}) - \theta(\BF{0})) \rangle
\label{eq:G}
\end{equation}
is predicted to behave at the critical temperature $T_c = 1/\beta_c$ 
as \cite{ami80}
\begin{equation}
G(\BF{x}) \propto \frac{(\ln |\BF{x}|)^{-2r}}{|\BF{x}|^\eta} 
\left[ 1 + {\cal O} \left( \frac{\ln \ln |\BF{x}|}{\ln |\BF{x}|} \right)\right],
\label{eq:G_crit}
\end{equation}
with $r = -1/16$ and $\eta = 1/4$. For the power of the logarithmic term we
have adopted the notation of Refs.~\cite{kenna96,kenna_plb}.
In the high-temperature phase near criticality, i.e. $0 < t \equiv T/T_c - 1 
\ll 1$, this implies for the magnetic susceptibility,
\begin{equation}
\chi = V \langle (\sum_{\BF{x}} \vec{s}(\BF{x})/V)^2 \rangle
     = \sum_{\BF{x}} G(\BF{x}),
\label{eq:chi}
\end{equation}
a scaling behavior
\begin{equation}
\chi \propto \xi^{2-\eta} (\ln \xi)^{-2r} 
[ 1 + {\cal O}( \ln \ln \xi/\ln \xi)],
\label{eq:chi_scal}
\end{equation}
where
\begin{equation}
\xi \propto \exp( b t^{-\nu})
\label{eq:xi}
\end{equation}
is the correlation length, with $\nu=1/2$ and $b$ being a non-universal
positive constant. Expressing $\xi$ in terms of $t$, eq.~(\ref{eq:chi_scal})
can also be written as 
\begin{equation}
\chi \propto \xi^{2-\eta} t^{2 \nu r}
[ ( 1 + {\cal O} ( t^\nu \ln t) ].
\label{eq:chi_scal2}
\end{equation}
Very close to $T_c$ eq.~(\ref{eq:chi_scal}) cannot hold for a finite 
system with linear size $L \ll \xi$. Here $\xi$ has to be replaced by
$L$, and we expect to observe a FSS behavior
\begin{equation}
\chi \propto L^{2-\eta} (\ln L)^{-2r}
[ 1 + {\cal O} ( \ln \ln L/\ln L )].
\label{eq:chi_fss}
\end{equation}

In numerical simulations it proved to be very difficult to
verify the KT scaling laws unambiguously. However, if one rejects
power-law ans\"atze with unnaturally large exponents and large confluent
correction terms, then, among the two alternatives, a pure power-law or
the exponential KT divergences, the KT predictions are clearly favored.
This is the conclusion of most numerical studies of the cosine formulation
\cite{wolff,gupta} and, with even stronger evidence, also of the Villain 
formulation \cite{ours} considered here.
In this note we shall therefore not study this fundamental question again.
We rather assume eqs.~(\ref{eq:chi_scal})-(\ref{eq:chi_fss}) to be 
qualitatively valid
and try to determine the exponents $\eta$, $\nu$, and $r$. Unfortunately even
this goal is far too ambitious, since a precise determination of all three 
critical exponents together with the (non-universal) value of $\beta_c$ would 
require much more accurate data than one can hope to generate with present
day techniques. We therefore hold the exponents $\nu=1/2$ and $\eta=1/4$
fixed at their theoretically predicted values and ask if any deviation of
the data from the leading scaling behavior can be explained by the logarithmic
corrections in eqs.~(\ref{eq:chi_scal}) and (\ref{eq:chi_fss}). 
%
                     \section{Results}
%
In Ref.~\cite{ours} we have reported high-precision MC simulations of the 
Villain model (\ref{eq:B_vil}), using the single-cluster update algorithm and
improved estimators for the two-point correlation function. This enabled us
to obtain on a $1200^2$ square lattice data for the correlation 
length up to $\xi \approx 140$. Since $L > 8 \xi$ this value should be a very
good approximation of
the thermodynamic limit. By performing fits of $\xi$ to the 
KT prediction (\ref{eq:xi}) and of $\chi$ to (\ref{eq:chi_scal}) (without
the logarithmic term) with four free parameters (the prefactor, $b$, $\nu$, and $\beta_c$) we obtained $\beta_c = 0.752(5)$ and $\nu=0.48(10)$. The estimate
of $\beta_c$ is in very good agreement with the more precise value of
$\beta_c = 0.7524(7)$ obtained in Ref.~\cite{dg_bc} from a study of the dual
discrete Gaussian model (see also \cite{dg_bc_old}). Using the ansatz 
(\ref{eq:chi_scal2}), i.e. including the theoretically predicted correction 
$t^{-1/16}$, did not improve the quality of the fits.

Further data of the susceptibility at criticality on lattices with
up to $512^2$ sites showed a clear scaling behavior for $L \ge 100$, 
$\chi \propto L^{2-\eta}$, with $\eta = 0.2495 \approx 1/4$ at $\beta=0.74$,
and $\eta = 0.2389(6) \ne 1/4$ at $\beta=0.75$. This is obviously not
consistent with the prediction that $\eta =1/4$ {\em at\/} $\beta_c$. Since
the estimate of $\beta_c$ from two completely independent simulations
agreed so well we concluded in Ref.~\cite{ours} that $\eta(\beta_c) \ne 1/4$,
in disagreement with the KT prediction. To reconcile simulations and theory we
speculated that the scaling curve for $\chi$ might still change for much
larger system sizes, but this is of course not very convincing. Mainly based
on our negative experience with the $t^{-1/16}$ correction in the $\chi(T)$
fits, we did not try, however, to attribute the observed discrepancy to 
logarithmic corrections.

The data at $\beta = 0.75$ and a fit in the range $L \ge 64$ 
according to $\ln(\chi/L^{7/4}) = {\rm const.} + (1/4-\eta) \ln L$ 
is reproduced in Fig.~1(a). In Fig.~1(b) we show
the same data, but now fix $\eta=1/4$ at the theoretical value and assume
that (\ref{eq:chi_fss}) {\em with} the logarithmic correction is valid.
Since then $\ln(\chi/L^{7/4}) = {\rm const.}- 2r \ln (\ln L)$, we expect a
straight line when $\ln(\chi/L^{7/4})$ is plotted against $\ln (\ln L)$. 
As is demonstrated in Fig.~1(b) this is clearly the case. Also shown is a 
linear fit which is of high statistical quality (goodness-of-fit parameter 
$Q=0.61$) and yields a slope of $0.0540(19)$, or 
\begin{equation}
r = -0.0270 \pm 0.0010,
\label{eq:r}
\end{equation}
in good agreement with the estimate of $r=-0.02(1)$ from the FSS of
Lee-Yang zeros in Refs.~\cite{kenna96,kenna_plb}. To summarize this subsection,
by allowing for logarithmic corrections we can reconcile the numerical
estimate of $\beta_c \approx 0.752$ with the KT prediction $\eta(\beta_c)=1/4$.
The value of the exponent $r$, however, is clearly {\em not} in agreement 
with the theoretical prediction $r = -1/16 = -0.0625$.

Let us next consider the scaling behavior of the thermodynamic data near
criticality in the high-temperature phase. In Ref.~\cite{ours} we neglected
logarithmic corrections in (\ref{eq:chi_scal}) and tested the relation
$\ln \chi/\xi^{7/4} = {\rm const.} + (1/4 - \eta) \ln \xi$ graphically. 
This plot
is reproduced in Fig.~2(a). We see that the curve has a negative slope,
corresponding to $\eta > 1/4$. We also observe, however, that the data are
curved and that for large $\xi$ the slope decreases. By defining 
$\eta^{\rm eff}$ from the local slopes, we obtained at the scale of 
$\xi \approx 110 \dots 140$ an estimate of $\eta^{\rm eff} \approx 0.267$.
Notice that this effective $\eta$ is {\em above} $1/4$, while from FSS
without logarithmic corrections we would have extracted an effective $\eta$
that is {\em smaller} than $1/4$. In Fig.~2(b) we show the same data, but
similar to Fig.~1 we now again fix $\eta=1/4$ at the theoretical value
and assume that (\ref{eq:chi_scal}) {\em with} the logarithmic correction
is valid. Since then $\ln(\chi/\xi^{7/4}) = {\rm const.} - 2r \ln (\ln \xi)$,
we expect a straight line when $\ln(\chi/\xi^{7/4})$ is plotted against 
$\ln (\ln \xi)$. This is indeed the case, and from the fit over all
available data points (with $Q=0.97$) we obtain
\begin{equation}
r = 0.0560 \pm 0.0017,
\label{eq:r_hight}
\end{equation}
in qualitative agreement with the results in Refs.~\cite{seiler,campostrini},
which are also derived from the approach to criticality in the 
high-temperature phase. The value (\ref{eq:r_hight}) is clearly different from
(\ref{eq:r}), and is very far from the theoretical estimate 
$r = -1/16 = - 0.0625$. In retrospective this ``explains'' why we did not
observe any improvement when trying fits of $\chi(T)$ {\em with} the
$t^r$ correction fixed to the theoretical prediction $t^{-1/16}$. 

We repeated the analysis leading to the Villain model estimate 
(\ref{eq:r_hight}) also with the three data
points for the cosine model in Ref.~\cite{ours} (with $\xi \approx 21$, 40, and
70) and obtained a compatible value of $r=0.047(8)$. Furthermore, using 
the more extensive data sets of Ref.~\cite{wolff} and Ref.~\cite{gupta} 
we find compatible values of $r=0.050(10)$ and $r=0.049(10)$, respectively.

We also tried to use the scaling form (\ref{eq:chi_scal2}) which
requires as input information the value of $\beta_c$. Using the most accurate
estimate of $\beta_c = 0.7524$ we find the result shown in Fig.~3. Again the
linear scaling looks almost perfect, but from the slope we now read off an
even larger value of $r = 0.0922(28)$. Qualitatively this can be understood
as follows. Going from (\ref{eq:chi_scal}) to (\ref{eq:chi_scal2}) we 
replace $\ln \xi$ by $t^{-\nu} = t^{-1/2}$. Asymptotically this follows from
the scaling behavior of $\xi$ in (\ref{eq:xi}). This implicitly assumes,
however, that the constant in the proper relation, 
$\ln \xi = {\rm const.} +bt^{-\nu}$, can be neglected. If $t$ is not really 
asymptotically small, this is not justified. In fact, the plot of  
$\ln(\ln \xi)$ vs $-\ln t$ in Fig.~4 does show effectively
an almost linear behavior, but with a slope completely different from the 
asymptotic value $\nu = 1/2$.

Finally it was of course tempting to enquire if the observed discrepancies
between the numerical data and the theoretical expectations can be blamed on
the additive logarithmic corrections in (\ref{eq:chi_scal}) and 
(\ref{eq:chi_fss}). To test this possibility we have replotted in Fig.~5
the data at criticality in the form
$\chi/L^{2-\eta}(\ln L)^{-2r}$ vs $\ln \ln L/\ln L$ and 
the thermodynamic data in the form
$\chi/\xi^{2-\eta}(\ln \xi)^{-2r}$ vs $\ln \ln \xi/\ln \xi$, assuming the
theoretically predicted values of $\eta$ and $r$. The double valuedness in
Fig.~5(b) is caused by the fact that $f(x) = \ln \ln x/\ln x$ assumes a
maximum $f_{\rm max} = 1/e \approx 0.3679$ at $x_{\rm max} = e^e \approx
15.15$. We see that both the data for $L > 64$ or $\xi > 40$ can be well 
fitted with a simple linear function. With a parabolic ansatz the acceptable
fit range can even be extended to smaller values of $L$ or $\xi$. From Fig.~5
it is obvious, however, that we are still too far away from the truly 
asymptotic region $x \longrightarrow 0$ to take this as a convincing 
evidence that additive logarithmic corrections can reconcile simulations and
theory.
%
                     \section{Discussion}
%
In summary we have shown that, when multiplicative logarithmic corrections 
are taken into account, numerical simulation data of the 2D XY Villain model
are quite consistent with the leading
KT predictions even at a quantitative level with critical exponents fixed to
the theoretical values of $\nu=1/2$ and $\eta=1/4$. Estimates of the 
logarithmic correction exponent $r$, however, turn out to be quite
inconsistent. Scaling analyses 
in the FSS region yield a negative ($r \approx -0.02 \dots -0.03$)
and analyses in the high-temparature phase a positive 
($r \approx 0.04 \dots 0.08$) value, both being quite different
from the theoretical prediction of $r = -1/16 = -0.0625$. This is obviously
related to the fact that analyses neglecting logarithmic corrections tended 
to estimate $\eta > 1/4$ using thermodynamic data and $\eta < 1/4$ in the
FSS region. We have no good explanation for this observation other than the
common, but unfortunately probably correct statement \cite{ggs82,cardy82}
that the studied system sizes are still much too small to resolve these 
discrepancies. 
%
                     \section*{Acknowledgements}
%
I would like to thank Ralph Kenna for inspiring discussions which initiated
this analysis, and Erhard Seiler for a copy of their updated paper prior
to publication. This work was supported by the DFG
through a Heisenberg fellowship which I gratefully acknowledge. 
%
%

\clearpage\newpage
\clearpage
%
%
\begin{figure}[htb]
\vskip 5.0truecm
\includegraphics{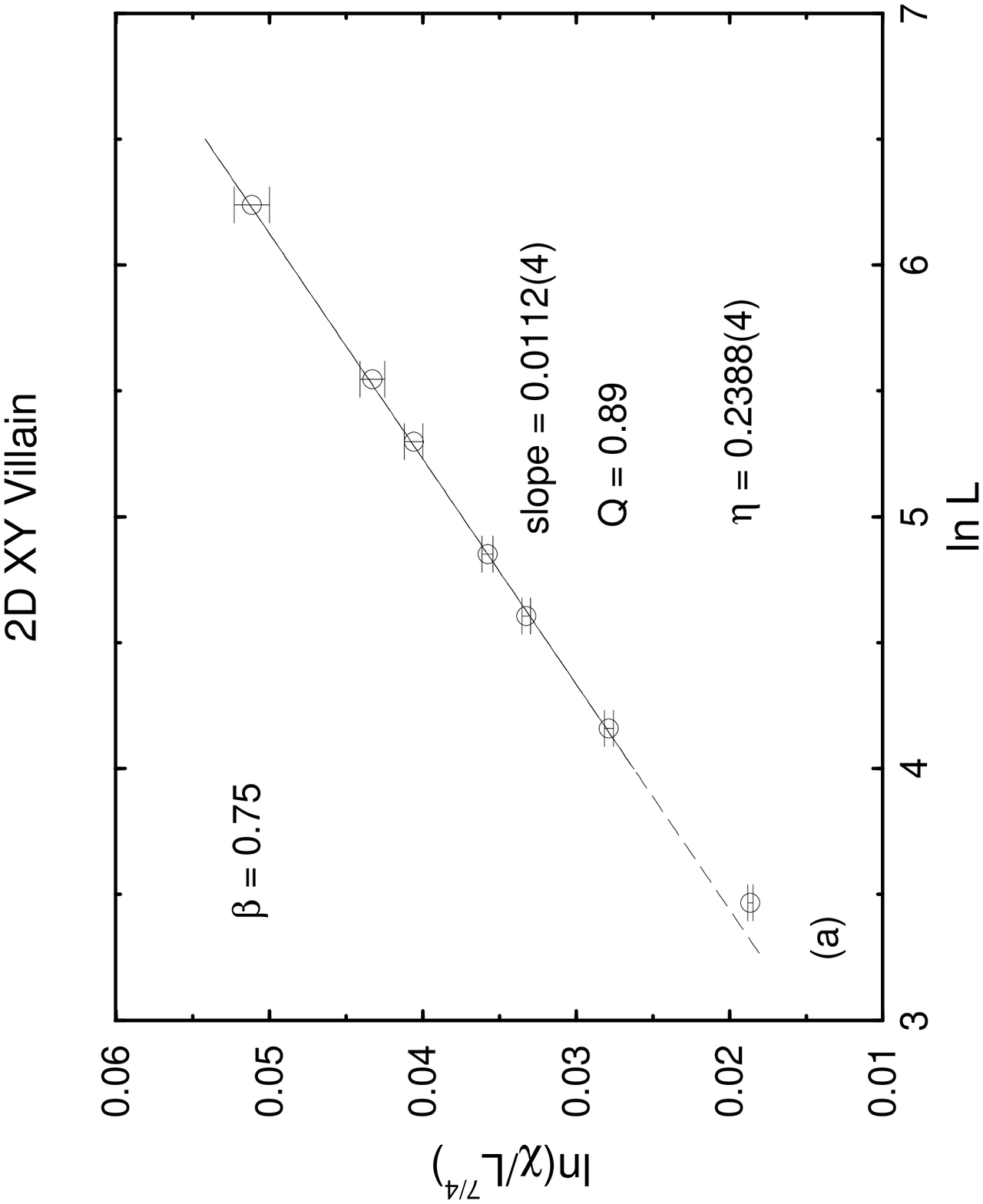}
\vskip 11.0truecm
\includegraphics{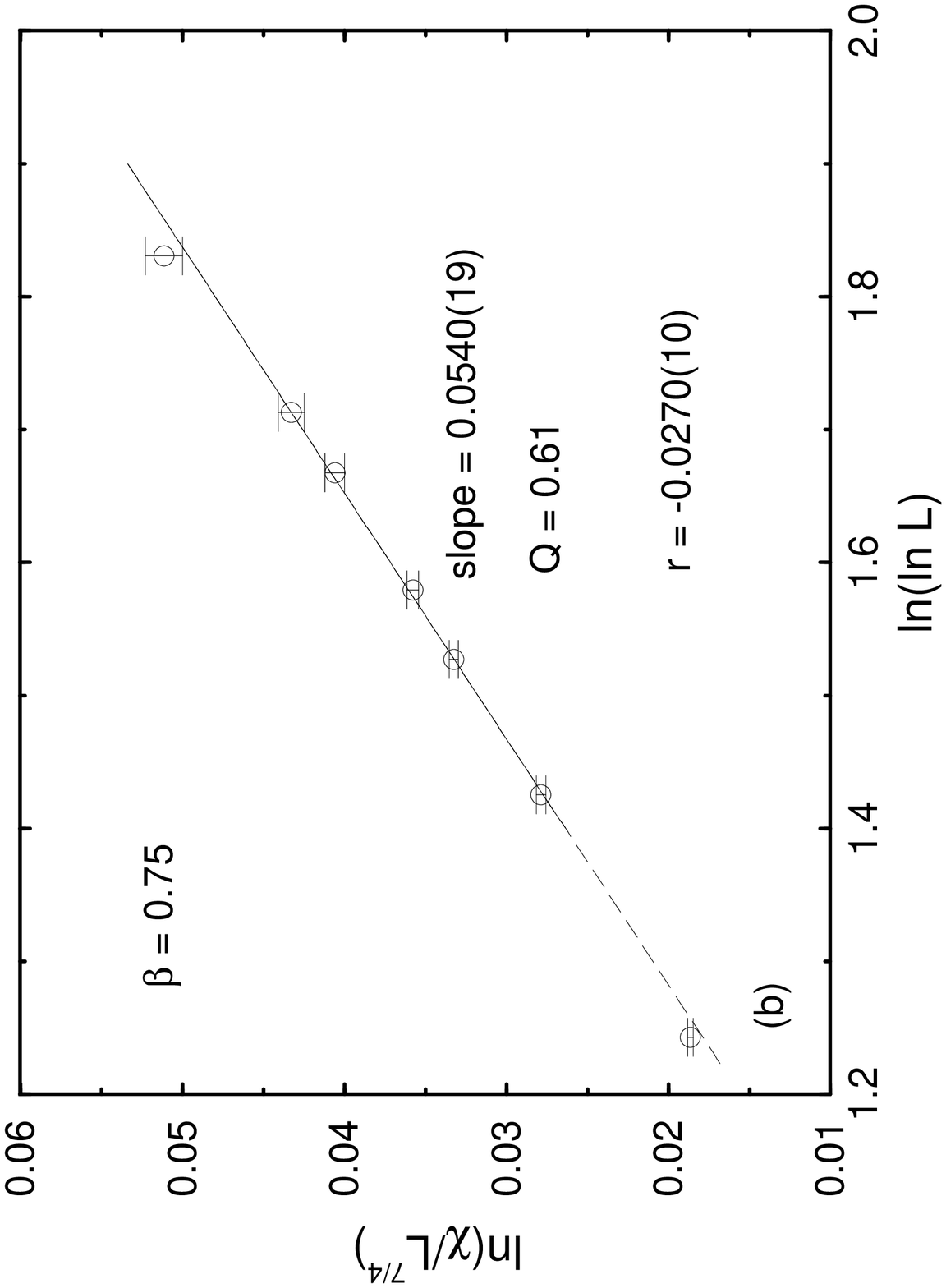}
\caption[]{\label{fig1} Finite-size scaling of the susceptibility at
criticality. If logarithmic corrections are neglected, the slope in
(a) gives $1/4 - \eta$. If $\eta=1/4$ is assumed, the slope in (b)
yields $-2r$, the exponent of the logarithmic correction.}
\end{figure}
%
%
\begin{figure}[htb]
\vskip 5.0truecm
\includegraphics{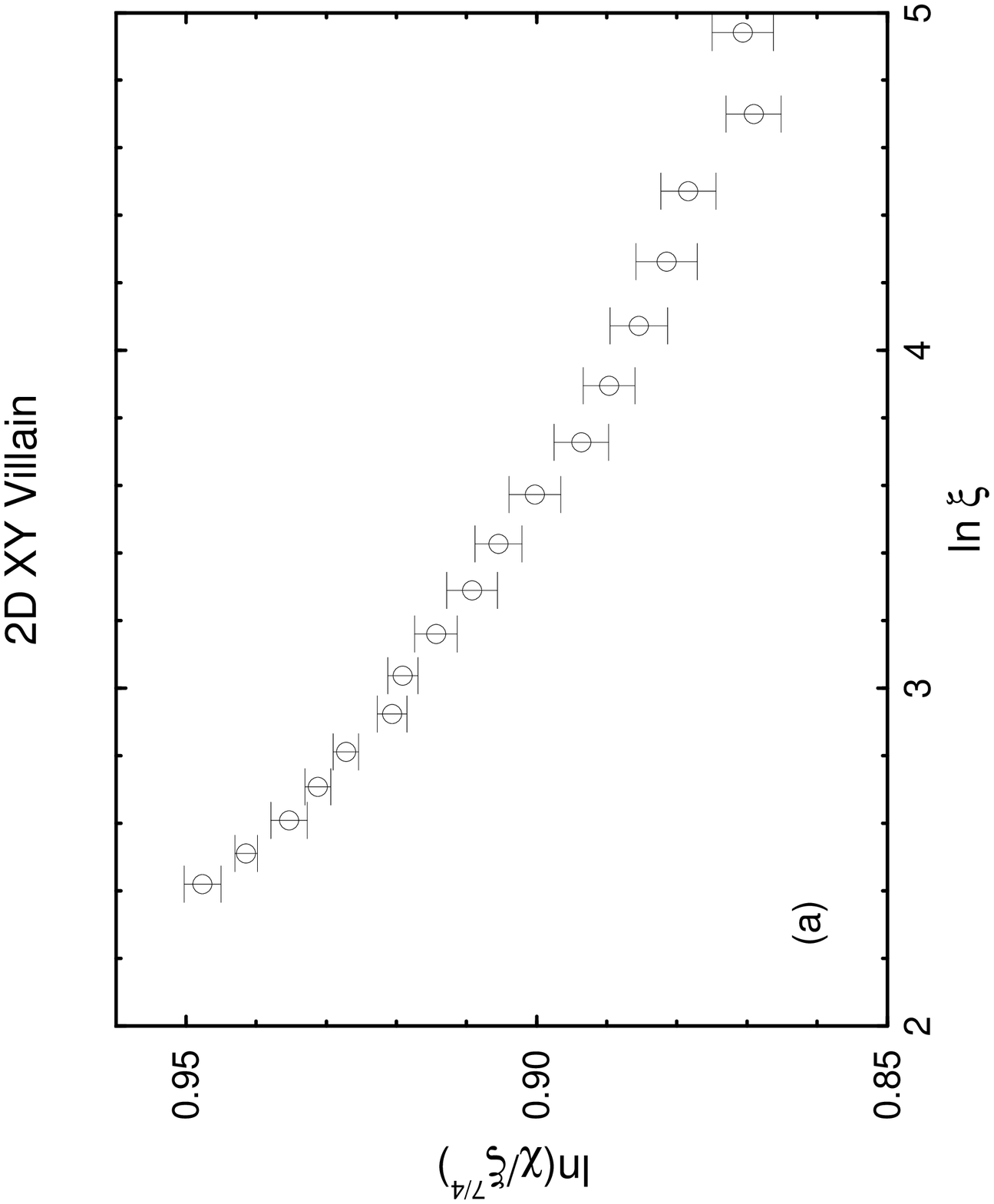}
\vskip 11.0truecm
\includegraphics{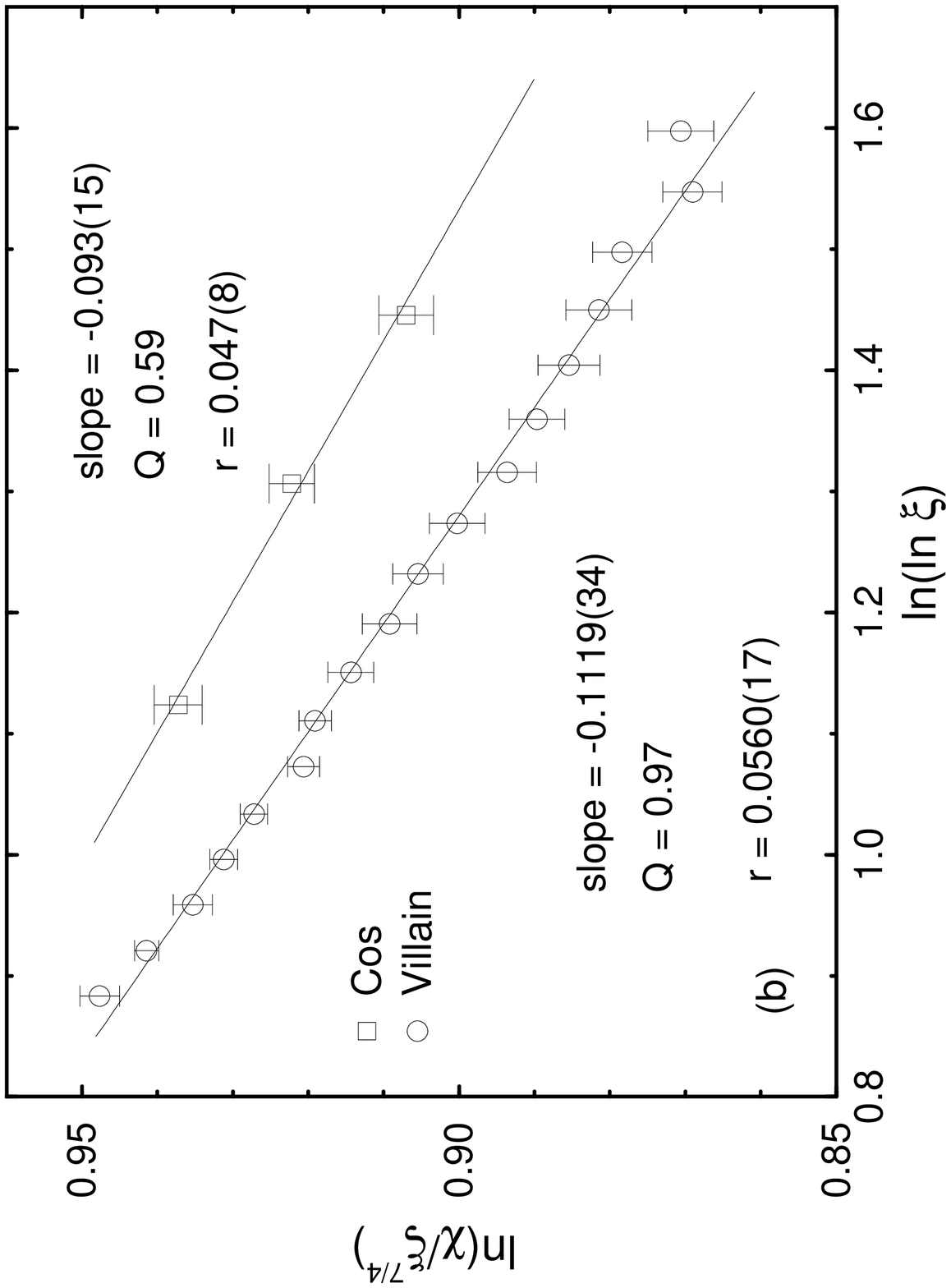}
\caption[]{\label{fig2} Test of the scaling relation $\chi \propto
\xi^{2-\eta} (\ln \xi)^{-2r}$ in the range $\xi \approx 10 \dots 140$, 
rewritten as $\ln(\chi/\xi^{7/4}) ={\rm const.} + (1/4-\eta) \ln \xi - 
2r \ln(\ln \xi)$. The linear behavior in (b) shows that the data are 
compatible with $\eta = 1/4$. As is already obvious from (a), the
exponent $r$ must then be positive, in disagreement with the theoretical
prediction $r = -1/16$.}
\end{figure}
%
%
\begin{figure}[htb]
\vskip 6.0truecm
\includegraphics{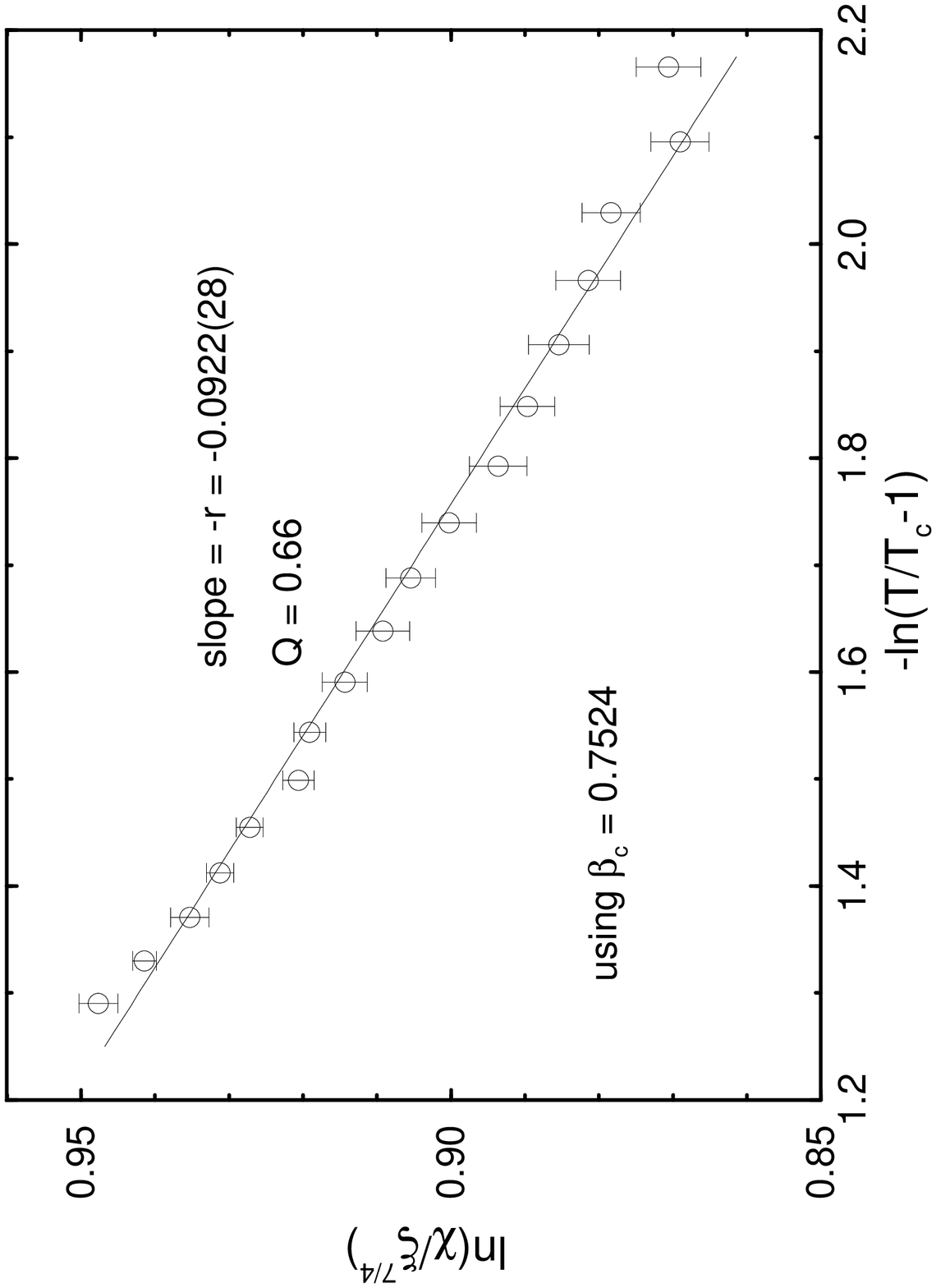}
\caption[]{\label{fig3} Test of scaling similar to Fig.~2, but with
$\ln \xi$ replaced by $t = T/T_c - 1$ (cp. eq.~(\ref{eq:xi})).}
\vskip 10.0truecm
\includegraphics{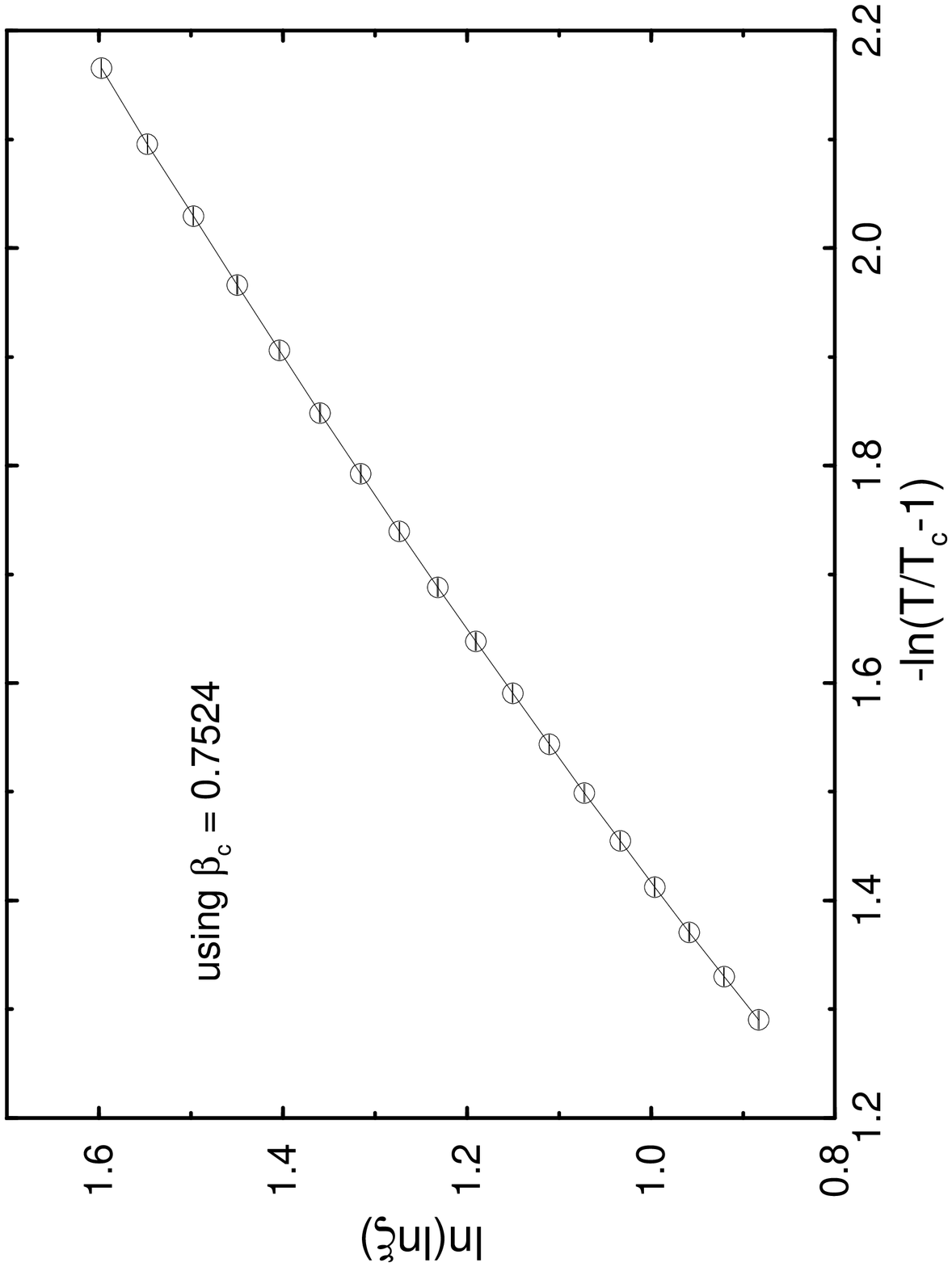}
\caption[]{\label{fig4} Correlation length vs reduced temperature.
In the range $\xi \approx 10 \dots 140$ the slope is effectively about 0.8, 
while asymptotically it should approach $\nu = 0.5$ according to
eq.~(\ref{eq:xi}).}  
\end{figure}
%
%
\begin{figure}[htb]
\vskip 5.0truecm
\includegraphics{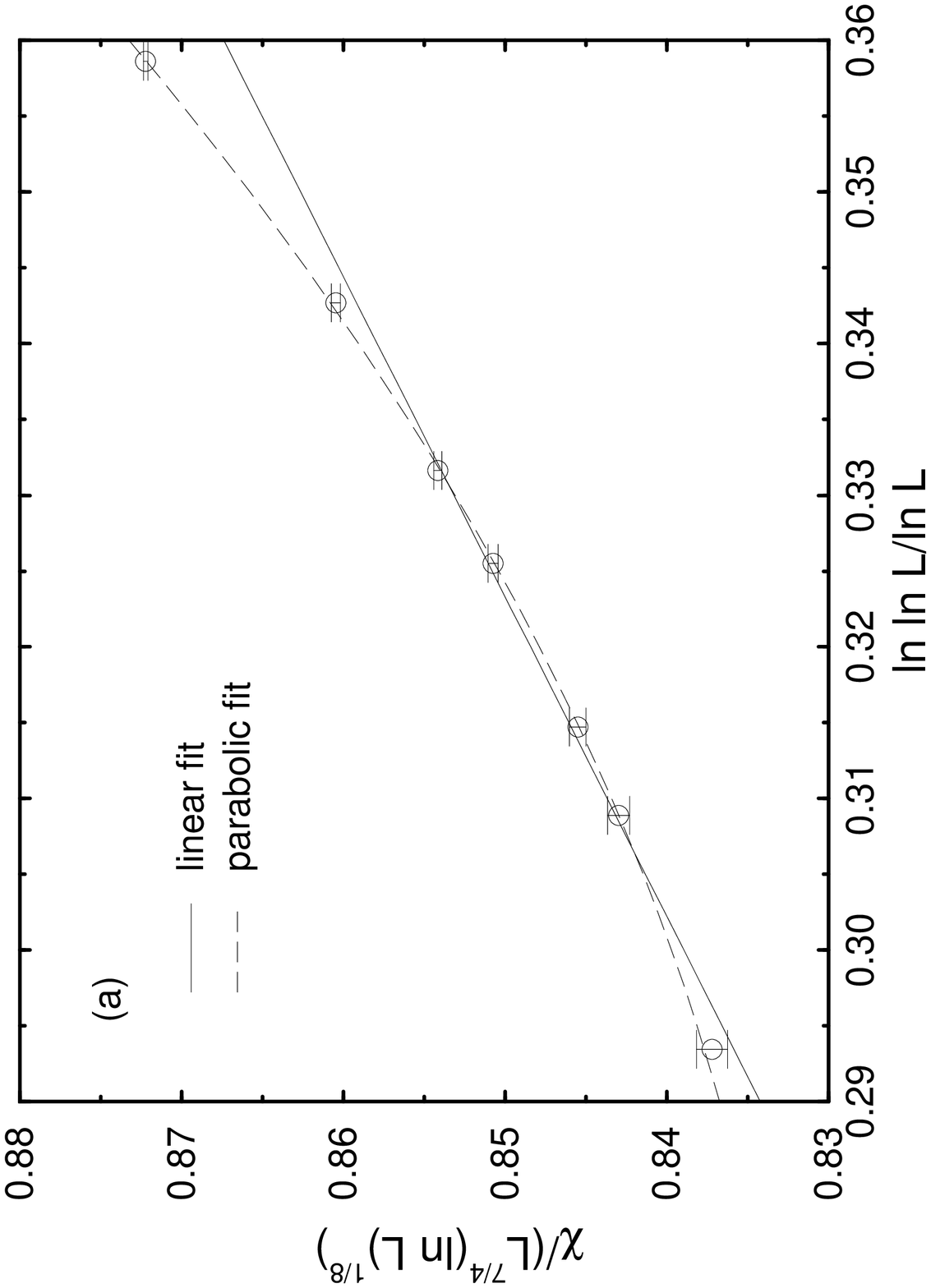}
\vskip 11.0truecm
\includegraphics{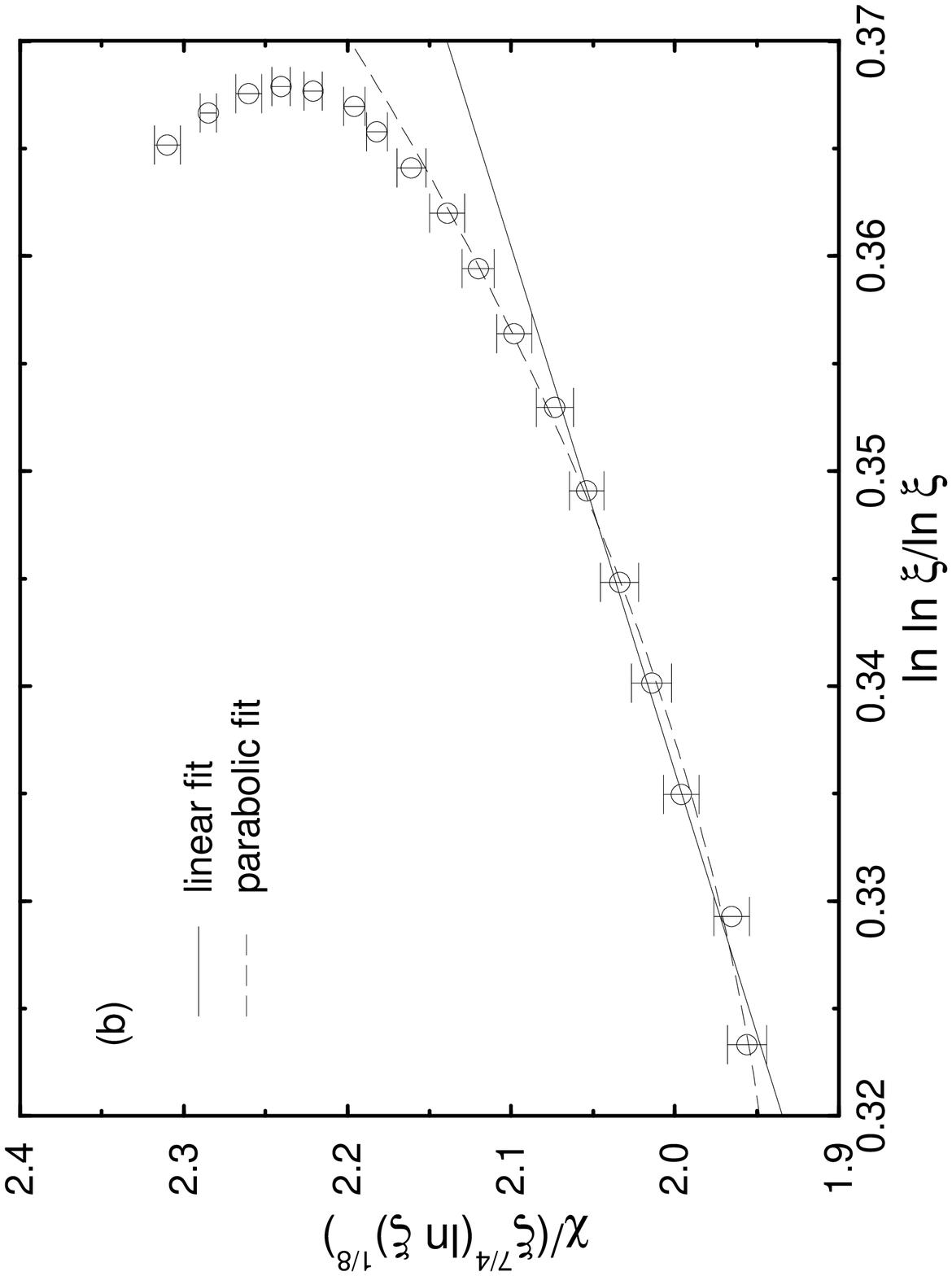}
\caption[]{\label{fig5} Test for additive logarithmic corrections
in (a) the data at criticality and (b) the thermodynamic data. Here
the exponents $\eta$ and $r$ are assumed to take the theoretically
predicted values $\eta=1/4$ and $r = -1/16$.}
\end{figure}
\end{document}